%% file: main.tex
\pdfoutput=1

\documentclass[12pt,a4paper]{article}

\usepackage{ifthen} 
\newboolean{pdflatex}
\setboolean{pdflatex}{true} 

\newboolean{articletitles}
\setboolean{articletitles}{true} 

\newboolean{uprightparticles}
\setboolean{uprightparticles}{false} 

\newboolean{inbibliography}
\setboolean{inbibliography}{false} 

\input{preamble}
\usepackage{longtable} 

\begin{document}

\renewcommand{\thefootnote}{\fnsymbol{footnote}}
\setcounter{footnote}{1}
\input{title-LHCb-PAPER}

\renewcommand{\thefootnote}{\arabic{footnote}}
\setcounter{footnote}{0}

\pagestyle{plain} 
\setcounter{page}{1}
\pagenumbering{arabic}


\input{paper}

\input{acknowledgements}

\addcontentsline{toc}{section}{References}
\setboolean{inbibliography}{true}
\bibliographystyle{LHCb}
\bibliography{main,references,LHCb-PAPER,LHCb-CONF,LHCb-DP,LHCb-TDR}

\newpage

\input{LHCb_Authorship_flat_16-Feb-2017.tex}

\end{document}

%% file: preamble.tex
\usepackage[top=1in, bottom=1.25in, left=1in, right=1in]{geometry}

\usepackage{microtype}
\usepackage{lineno}  
\usepackage{xspace} 
\usepackage{caption} 

\usepackage{graphicx}  
\usepackage{color}
\usepackage{colortbl}
\graphicspath{{./figs/}} 

\usepackage{amsmath} 
\usepackage{amssymb}
\usepackage{amsfonts}
\usepackage{upgreek} 

\newcommand*\patchAmsMathEnvironmentForLineno[1]{%
\expandafter\let\csname old#1\expandafter\endcsname\csname #1\endcsname
\expandafter\let\csname oldend#1\expandafter\endcsname\csname
end#1\endcsname
 \renewenvironment{#1}%
   {\linenomath\csname old#1\endcsname}%
   {\csname oldend#1\endcsname\endlinenomath}%
}
\newcommand*\patchBothAmsMathEnvironmentsForLineno[1]{%
  \patchAmsMathEnvironmentForLineno{#1}%
  \patchAmsMathEnvironmentForLineno{#1*}%
}
\AtBeginDocument{%
\patchBothAmsMathEnvironmentsForLineno{equation}%
\patchBothAmsMathEnvironmentsForLineno{align}%
\patchBothAmsMathEnvironmentsForLineno{flalign}%
\patchBothAmsMathEnvironmentsForLineno{alignat}%
\patchBothAmsMathEnvironmentsForLineno{gather}%
\patchBothAmsMathEnvironmentsForLineno{multline}%
\patchBothAmsMathEnvironmentsForLineno{eqnarray}%
}

\usepackage{hyperref}    
\usepackage[all]{hypcap} 

\input{lhcb-symbols-def} 

\usepackage{cite} 
\usepackage{mciteplus}

%% file: lhcb-symbols-def.tex
\usepackage{xspace} 
\usepackage{upgreek}

\def\lhcb {\mbox{LHCb}\xspace}

\def\MagUp {\mbox{\em Mag\kern -0.05em Up}\xspace}

\ifthenelse{\boolean{uprightparticles}}%
{

 \def\Ppi         {\ensuremath{\uppi}\xspace}

 \def\PDelta      {\ensuremath{\Delta}\xspace}                 
 \def\PXi      {\ensuremath{\Xi}\xspace}                 
 \def\PLambda      {\ensuremath{\Lambda}\xspace}                 
 \def\PSigma      {\ensuremath{\Sigma}\xspace}                 
 \def\POmega      {\ensuremath{\Omega}\xspace}                 
 \def\PUpsilon      {\ensuremath{\Upsilon}\xspace}                 
 
 \def\PB      {\ensuremath{\mathrm{B}}\xspace}                 
                  
 \def\PD      {\ensuremath{\mathrm{D}}\xspace}

 \def\PK      {\ensuremath{\mathrm{K}}\xspace}

 \def\Pb      {\ensuremath{\mathrm{b}}\xspace}                 
 \def\Pc      {\ensuremath{\mathrm{c}}\xspace}                 
                  
 \def\Pe      {\ensuremath{\mathrm{e}}\xspace}

 \def\Pi      {\ensuremath{\mathrm{i}}\xspace}

 \def\Ps      {\ensuremath{\mathrm{s}}\xspace}

}
{

 \def\Ppi         {\ensuremath{\pi}\xspace}

 \mathchardef\PDelta="7101
 \mathchardef\PXi="7104
 \mathchardef\PLambda="7103
 \mathchardef\PSigma="7106
 \mathchardef\POmega="710A
 \mathchardef\PUpsilon="7107
                  
 \def\PB      {\ensuremath{B}\xspace}                 
                  
 \def\PD      {\ensuremath{D}\xspace}

 \def\PK      {\ensuremath{K}\xspace}

 \def\Pb      {\ensuremath{b}\xspace}                 
 \def\Pc      {\ensuremath{c}\xspace}                 
                  
 \def\Pe      {\ensuremath{e}\xspace}

 \def\Pi      {\ensuremath{i}\xspace}

 \def\Ps      {\ensuremath{s}\xspace}

}

\makeatletter
\ifcase \@ptsize \relax
  \newcommand{\miniscule}{\@setfontsize\miniscule{4}{5}}
\or
  \newcommand{\miniscule}{\@setfontsize\miniscule{5}{6}}
\or
  \newcommand{\miniscule}{\@setfontsize\miniscule{5}{6}}
\fi
\makeatother

\DeclareRobustCommand{\optbar}[1]{\shortstack{{\miniscule (\rule[.5ex]{1.25em}{.18mm})}
  \\ [-.7ex] $#1$}}

\def\epem       {{\ensuremath{\Pe^+\Pe^-}}\xspace}

\def\squark    {{\ensuremath{\Ps}}\xspace}

\def\cquark    {{\ensuremath{\Pc}}\xspace}

\def\bquark    {{\ensuremath{\Pb}}\xspace}

\def\pion   {{\ensuremath{\Ppi}}\xspace}
\def\piz    {{\ensuremath{\pion^0}}\xspace}

\def\pip    {{\ensuremath{\pion^+}}\xspace}
\def\pim    {{\ensuremath{\pion^-}}\xspace}

\def\pimp   {{\ensuremath{\pion^\mp}}\xspace}

\def\kaon    {{\ensuremath{\PK}}\xspace}
  \def\Kbar    {{\kern 0.2em\overline{\kern -0.2em \PK}{}}\xspace}

\def\KorKbar    {\kern 0.18em\optbar{\kern -0.18em K}{}\xspace}

\def\Kp      {{\ensuremath{\kaon^+}}\xspace}
\def\Km      {{\ensuremath{\kaon^-}}\xspace}

  \def\Dbar    {{\kern 0.2em\overline{\kern -0.2em \PD}{}}\xspace}
\def\D       {{\ensuremath{\PD}}\xspace}

\def\DorDbar    {\kern 0.18em\optbar{\kern -0.18em D}{}\xspace}

\def\Dzb     {{\ensuremath{\Dbar{}^0}}\xspace}

\def\Dm      {{\ensuremath{\D^-}}\xspace}
\def\Dpm     {{\ensuremath{\D^\pm}}\xspace}

\def\Dstar   {{\ensuremath{\D^*}}\xspace}

\def\Dstarm  {{\ensuremath{\D^{*-}}}\xspace}

\def\DorDstar   {{\ensuremath{\D^{(*)}}}\xspace}

\def\DorDstarp  {{\ensuremath{\D^{(*)+}}}\xspace}
\def\DorDstarm  {{\ensuremath{\D^{(*)-}}}\xspace}
\def\DorDstarpm {{\ensuremath{\D^{(*)\pm}}}\xspace}

\def\B       {{\ensuremath{\PB}}\xspace}
\def\Bbar    {{\ensuremath{\kern 0.18em\overline{\kern -0.18em \PB}{}}}\xspace}

\def\BorBbar    {\kern 0.18em\optbar{\kern -0.18em B}{}\xspace}
\def\Bz      {{\ensuremath{\B^0}}\xspace}
\def\Bzb     {{\ensuremath{\Bbar{}^0}}\xspace}
\def\Bu      {{\ensuremath{\B^+}}\xspace}
\def\Bub     {{\ensuremath{\B^-}}\xspace}
\def\Bp      {{\ensuremath{\Bu}}\xspace}
\def\Bm      {{\ensuremath{\Bub}}\xspace}

\def\Bs      {{\ensuremath{\B^0_\squark}}\xspace}

  \def\Y#1S{\ensuremath{\PUpsilon{(#1S)}}\xspace}

\def\FourS {{\Y4S}}

\def\Lbar        {{\ensuremath{\kern 0.1em\overline{\kern -0.1em\PLambda}}}\xspace}
\def\LorLbar    {\kern 0.18em\optbar{\kern -0.18em \PLambda}{}\xspace}

\def\to                 {\ensuremath{\rightarrow}\xspace}

\def\AT#1     {\ensuremath{A_{\mathrm{T}}^{#1}}\xspace}           

\def\C#1      {\ensuremath{\mathcal{C}_{#1}}\xspace}                       
\def\Cp#1     {\ensuremath{\mathcal{C}_{#1}^{'}}\xspace}                    
\def\Ceff#1   {\ensuremath{\mathcal{C}_{#1}^{\mathrm{(eff)}}}\xspace}        
\def\Cpeff#1  {\ensuremath{\mathcal{C}_{#1}^{'\mathrm{(eff)}}}\xspace}       
\def\Ope#1    {\ensuremath{\mathcal{O}_{#1}}\xspace}                       
\def\Opep#1   {\ensuremath{\mathcal{O}_{#1}^{'}}\xspace}                    

\newcommand{\tev}{\ifthenelse{\boolean{inbibliography}}{\ensuremath{~T\kern -0.05em eV}}{\ensuremath{\mathrm{\,Te\kern -0.1em V}}}\xspace}
\newcommand{\gev}{\ensuremath{\mathrm{\,Ge\kern -0.1em V}}\xspace}
\newcommand{\mev}{\ensuremath{\mathrm{\,Me\kern -0.1em V}}\xspace}
\newcommand{\kev}{\ensuremath{\mathrm{\,ke\kern -0.1em V}}\xspace}
\newcommand{\ev}{\ensuremath{\mathrm{\,e\kern -0.1em V}}\xspace}
\newcommand{\nspgevc}{\ensuremath{{\mathrm{Ge\kern -0.1em V\!/}c}}\xspace}
\newcommand{\gevc}{\ensuremath{{\mathrm{\,Ge\kern -0.1em V\!/}c}}\xspace}
\newcommand{\mevc}{\ensuremath{{\mathrm{\,Me\kern -0.1em V\!/}c}}\xspace}
\newcommand{\gevcc}{\ensuremath{{\mathrm{\,Ge\kern -0.1em V\!/}c^2}}\xspace}
\newcommand{\gevgevcccc}{\ensuremath{{\mathrm{\,Ge\kern -0.1em V^2\!/}c^4}}\xspace}
\newcommand{\mevcc}{\ensuremath{{\mathrm{\,Me\kern -0.1em V\!/}c^2}}\xspace}

\def\mum  {\ensuremath{{\,\upmu\mathrm{m}}}\xspace}

\def\invfb   {\ensuremath{\mbox{\,fb}^{-1}}\xspace}

\newcommand{\chisq}{\ensuremath{\chi^2}\xspace}

\newcommand{\chisqip}{\ensuremath{\chi^2_{\text{IP}}}\xspace}

\newcommand{\chisqvtx}{\ensuremath{\chi^2_{\text{vtx}}}\xspace}

\def\gsim{{~\raise.15em\hbox{$>$}\kern-.85em
          \lower.35em\hbox{$\sim$}~}\xspace}
\def\lsim{{~\raise.15em\hbox{$<$}\kern-.85em
          \lower.35em\hbox{$\sim$}~}\xspace}

\def\sPlot{\mbox{\em sPlot}\xspace}

\def\ptot       {\mbox{$p$}\xspace}
\def\pt         {\mbox{$p_{\mathrm{ T}}$}\xspace}

\def\evtgen     {\mbox{\textsc{EvtGen}}\xspace}

\def\geant      {\mbox{\textsc{Geant4}}\xspace}

\def\photos     {\mbox{\textsc{Photos}}\xspace}

\def\pythia     {\mbox{\textsc{Pythia}}\xspace}

\def\tell1  {TELL1\xspace}
\def\ukl1   {UKL1\xspace}

\newcommand{\ie}{\mbox{\itshape i.e.}\xspace}

%% file: title-LHCb-PAPER.tex
\begin{titlepage}
\pagenumbering{roman}

\vspace*{-1.5cm}
\centerline{\large EUROPEAN ORGANIZATION FOR NUCLEAR RESEARCH (CERN)}
\vspace*{1.5cm}
\noindent
\begin{tabular*}{\linewidth}{lc@{\extracolsep{\fill}}r@{\extracolsep{0pt}}}
\vspace*{-2.7cm}\mbox{\!\!\!\includegraphics[width=.14\textwidth]{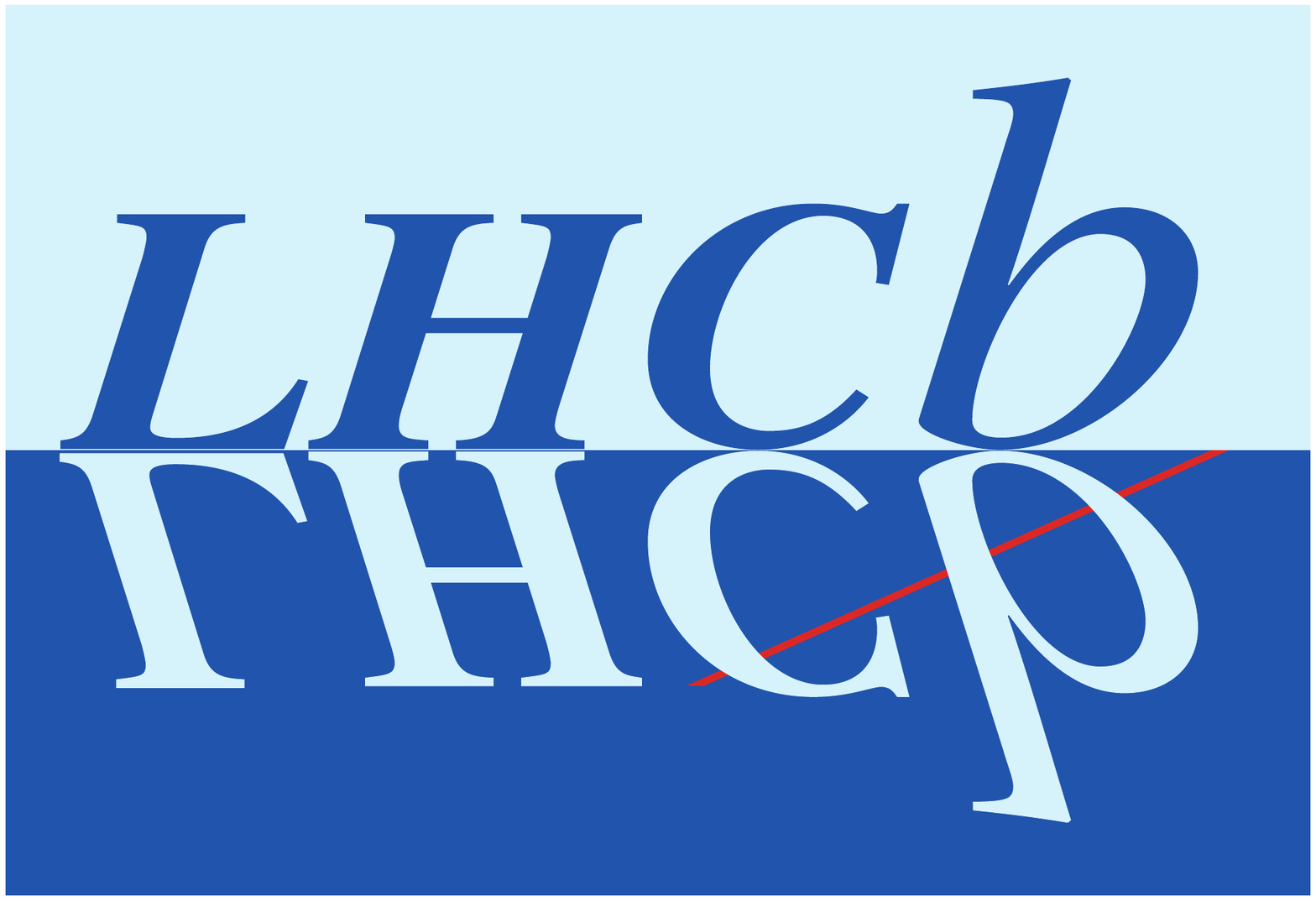}} & & \\
 & & CERN-EP-2017-064 \\  
 & & LHCb-PAPER-2017-006 \\  
 & & \today \\ 
 & & \\
\end{tabular*}

\vspace*{4.0cm}

{\normalfont\bfseries\boldmath\huge
\begin{center}
  Observation of the $\Bp \to \Dstarm\Kp\pip$ decay
\end{center}
}

\vspace*{2.0cm}

\begin{center}
  The LHCb collaboration\footnote{Authors are listed at the end of this paper.}
\end{center}

\vspace{\fill}

\begin{abstract}
  \noindent
  The $\Bp \to \Dstarm\Kp\pip$ decay potentially provides an excellent way to investigate charm meson spectroscopy.
  The decay is searched for in a sample of proton-proton collision data
  collected with the LHCb detector at centre-of-mass energies of 7 and
  8\tev, corresponding to an integrated luminosity of 3\invfb.
  A clear signal is observed,
  and the ratio of its branching fraction to that of the $\Bp \to \Dstarm \pip \pip$ normalisation channel is measured to be
  \begin{equation*}
    \frac{ \mathcal{B} (\Bp \to \Dstarm \Kp \pip ) }{\mathcal{B} ( \Bp \to \Dstarm \pip \pip )} = \left( 6.39 \pm 0.27 \pm 0.48 \right) \times 10^{-2} \, ,
  \end{equation*}
  where the first uncertainty is statistical and the second is systematic.
  This is the first observation of the $\Bp \to \Dstarm\Kp\pip$ decay.
\end{abstract}

\vspace*{2.0cm}

\begin{center}
  Published as \href{http://dx.doi.org/10.1103/PhysRevD.96.011101}{Phys.~Rev.~{\bf D96}~(2017)~011101(R)}
\end{center}

\vspace{\fill}

{\footnotesize 
\centerline{\copyright~CERN on behalf of the \lhcb collaboration, licence \href{http://creativecommons.org/licenses/by/4.0/}{CC-BY-4.0}.}}
\vspace*{2mm}

\end{titlepage}

\newpage
\setcounter{page}{2}
\mbox{~}

%% file: paper.tex
The $\B \to \DorDstar h h^{\prime}$ decays, where $h^{(\prime)} = \pi, K$, provide an excellent way to investigate the spectroscopy of excited charm mesons.
The constrained initial and final states lead to comparatively low backgrounds and excellent mass resolution, and furthermore amplitude analysis can be used to determine the quantum numbers of any intermediate resonant states through their angular distributions. 
By contrast, very large yields are available through inclusive production of excited charm states, but studies of such processes cannot in general result in unambiguous determinations of quantum numbers, and the sizable backgrounds tend to lead to large systematic uncertainties.

The amplitude analysis approach has been pursued extensively for $\B \to D h h^{\prime}$ decays.
For the cases where both $hh^{\prime}$ particles are pions, the $\Bp \to \Dm\pip\pip$ and $\Bz \to \Dzb \pip\pim$ decays have been studied by the Belle~\cite{Abe:2003zm,Kuzmin:2006mw}, BaBar~\cite{Aubert:2009wg,:2010ip} and LHCb~\cite{LHCb-PAPER-2014-070,LHCb-PAPER-2016-026} collaborations.
Regarding modes with a kaon in the final state, detailed analyses of $\Bp \to \Dm\Kp\pip$~\cite{LHCb-PAPER-2015-007}, $\Bz \to \Dzb \Kp\pim$~\cite{LHCb-PAPER-2015-017} and $\Bs \to \Dzb \Km\pip$~\cite{LHCb-PAPER-2014-035,LHCb-PAPER-2014-036} decays have been performed by LHCb.
In spite of the Cabibbo suppression of the $\Bp$ and $\Bz$ decays to the final states containing kaons compared to those with only pions, sufficiently large samples can be obtained to provide useful independent measurements of the properties of excited charm mesons.

The above-mentioned decays are, however, only sensitive to resonant states with natural spin-parity, \ie\ with $J^P$ in the series $0^+, 1^-, 2^+, 3^-, ...$, as only those states can decay strongly to two pseudoscalar mesons.
Relatively little information exists on the states with unnatural spin-parity.
Apart from work on the $\Bp \to \Dstarm\pip\pip$ mode by Belle~\cite{Abe:2003zm}, there has been no experimental study of the resonant substructure of $\B \to \Dstar h h^{\prime}$ decays.
Studies of inclusive production of $\Dstar\pi$ resonances in $\epem$ and $pp$ collisions have been made by BaBar~\cite{delAmoSanchez:2010vq} and LHCb~\cite{LHCb-PAPER-2013-026}, respectively, but more detailed investigations are necessary to understand the spectrum of states.

Decays of the form $\B \to \DorDstar K\pi$ are also important in the context of determining the angle $\gamma$ of the Cabibbo-Kobayashi-Maskawa (CKM) quark mixing matrix~\cite{Cabibbo:1963yz,Kobayashi:1973fv}.
Sensitivity to $\gamma$ arises when amplitudes proportional to the CKM matrix elements $V^{}_{ub}V^{*}_{cs}$ and $V^{}_{cb}V^{*}_{us}$ interfere, and so the $\DorDstar$ meson must decay into a final state accessible to both charm flavour eigenstates.
This is not possible for $\DorDstarp$ decays.
However, the relative rates of $\Bp \to \DorDstarp \Kp \pim$ and $\Bp \to \DorDstarm \Kp \pip$ decays through an intermediate $\DorDstarpm \pimp$ resonance can be used to determine the relative magnitude of the two amplitudes~\cite{Sinha:2004ct}, as was recently done for $\Bp \to \Dpm \Kp \pimp$ decays~\cite{LHCb-PAPER-2015-007,LHCb-PAPER-2015-054}.
This information may subsequently be used as an external constraint in a determination of $\gamma$ from decays of the same resonance in the $\Bp \to \DorDstar \Kp \piz$ final state.
Moreover, as an increasingly wide range of decays are being used to obtain constraints on $\gamma$~\cite{LHCb-PAPER-2016-032}, it is important to improve knowledge of modes such as $\B \to \Dstar h h^{\prime}$ which may cause backgrounds and hence systematic uncertainties in the analyses.

\begin{figure}[!tb]
  \centering
  \includegraphics[width=0.55\textwidth]{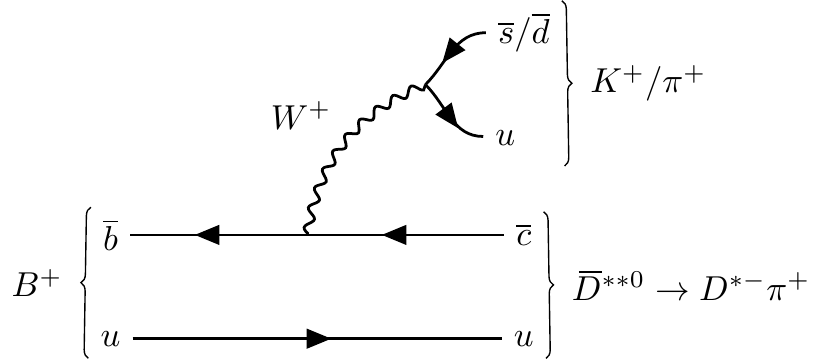}
  \caption{\small
    Leading diagram for $\Bp \to \Dstarm \Kp \pip$ and $\Dstarm \pip\pip$ decays,
    where the $\Dstarm \pip$ system is produced through the decay of an
    excited charm state denoted $\Dbar{}^{**0}$.
  }
  \label{fig:feynman}
\end{figure}

In this paper, the first search for the $\Bp \to \Dstarm \Kp \pip$ decay is presented.
The $\Dstarm$ meson is reconstructed through its decay to $\Dzb\pim$ with $\Dzb\to\Kp\pim$.
The topologically similar $\Bp \to \Dstarm \pip\pip$ decay is used as a control channel and for normalisation of the branching fraction measurement.
The leading diagram for $\Bp \to \Dstarm \Kp \pip$ and $\Dstarm \pip\pip$ decays is shown in Fig.~\ref{fig:feynman}.
The inclusion of charge-conjugate processes is implied throughout the paper.
The analysis is based on procedures used for previous analyses of similar decay modes~\cite{LHCb-PAPER-2015-007,LHCb-PAPER-2016-026}.
An important feature is that signal decays have a narrow peak in the distribution of $\Delta m$, the difference between the $\Dstarm$ and $\Dzb$ candidate masses; imposing a requirement on $\Delta m$ greatly reduces the range of possible sources of background.

The analysis is based on a sample of proton-proton collision data collected
with the LHCb detector at centre-of-mass energies of 7 and 8\tev, corresponding
to an integrated luminosity of 3\invfb.
The LHCb detector~\cite{Alves:2008zz,LHCb-DP-2014-002} is a single-arm forward spectrometer covering the \mbox{pseudorapidity} range $2<\eta <5$, designed for the study of particles containing \bquark or \cquark quarks. 
The detector includes a high-precision tracking system consisting of a silicon-strip vertex detector surrounding the $pp$ interaction region, a large-area silicon-strip detector located upstream of a dipole magnet with a bending power of about $4\,{\rm Tm}$ and three stations of silicon-strip detectors and straw drift tubes placed downstream of the magnet.
The tracking system provides a measurement of momentum, \ptot, of charged particles with relative uncertainty that varies from $0.5\,\%$ at low momentum to $1.0\,\%$ at $200\gevc$.
The minimum distance of a track to a primary vertex (PV), the impact parameter (IP), is measured with a resolution of $(15+29/\pt)\mum$,
where \pt is the component of the momentum transverse to the beam, in $\nspgevc$.
Different types of charged hadrons are distinguished using information from two ring-imaging Cherenkov detectors. 
Photons, electrons and hadrons are identified by a calorimeter system consisting of scintillating-pad and preshower detectors, an electromagnetic calorimeter and a hadronic calorimeter. 
Muons are identified by a system composed of alternating layers of iron and multiwire proportional chambers.

Online event selection is performed by a trigger, which consists of a hardware stage, based on information from the calorimeter and muon systems, followed by a software stage,  in which all tracks with $\pt>500~(300)\mevc$ are reconstructed for data collected in 2011 (2012).
At the hardware trigger stage, events are required to contain either a muon with high transverse momentum or a particle that deposits high transverse energy in the calorimeters.
For hadrons, the transverse energy threshold is typically $3.5 \gev$.
The software trigger used in the analysis reported in this paper requires a two-, three- or four-track secondary vertex with significant displacement from any PV. 
At least one charged particle must have \pt\ above a threshold of $1.7\,(1.6) \gevc$ in the $\sqrt{s} = 7\,(8) \tev$ data. 
This particle must also be inconsistent with originating from any PV, as quantified through the difference in the vertex fit \chisq of a given PV reconstructed with and without the considered particle (\chisqip).
A multivariate algorithm~\cite{BBDT} is used for the identification of secondary vertices consistent with the decay of a \bquark hadron.
In the offline selection, the objects that fired the trigger are associated with reconstructed particles.  
Selection requirements can therefore be made not only on the particular trigger that fired, but on whether the decision was due to the signal candidate, other particles produced in the $pp$ collision, or a combination of both~\cite{LHCb-DP-2012-004}.
Candidates are retained from events in which the hardware trigger is caused either by the signal candidate or by other particles in the event.
In the former case, it is further required that the trigger is caused by the deposits of the signal decay products in the calorimeters.
It is also required that the software trigger decision must have been caused entirely by tracks that form the signal candidate.

Simulated events are used to characterise the detector response to signal and certain types of background events.
In the simulation, $pp$ collisions are generated using \pythia~\cite{Sjostrand:2006za,*Sjostrand:2007gs} with a specific \lhcb configuration~\cite{LHCb-PROC-2010-056}.  
Decays of hadronic particles are described by \evtgen~\cite{Lange:2001uf}, in which final-state radiation is generated using \photos~\cite{Golonka:2005pn}. 
The interaction of the generated particles with the detector and its response are implemented using the \geant toolkit~\cite{Allison:2006ve, *Agostinelli:2002hh} as described in Ref.~\cite{LHCb-PROC-2011-006}.

Candidates consistent with the decay chains $\Bp \to \Dstarm \Kp \pip$ and $\Bp \to \Dstarm \pip\pip$, with $\Dstarm \to \Dzb\pim$ and $\Dzb\to\Kp\pim$, are selected.
The criteria for $\Bp \to \Dstarm \Kp \pip$ and $\Bp \to \Dstarm \pip\pip$ candidates are identical, except for particle identification requirements (discussed below).
Loose initial selection requirements on the quality of the tracks combined to form the \Bp\ candidate, as well as on their $p$, $\pt$ and $\chisqip$, are applied.
The \Dzb\ candidate must have invariant mass within $\pm 100 \mevcc$ of the known \Dzb\ mass~\cite{PDG2016}.
Further requirements are imposed on the vertex quality ($\chisqvtx$) and flight distance of the \Bp\ and \Dzb\ candidates from the PV with which they have the smallest \chisqip (for the \Bp\ candidate, this is referred to as the associated PV).
The \Bp\ candidate must also satisfy requirements on its invariant mass and on the cosine of the angle between the momentum vector and the line joining the \B\ vertex to the associated PV.
The value of $\Delta m$ is required to be less than $5 \mevcc$ from the known difference between the $\Dstarm$ and $\Dzb$ masses~\cite{PDG2016}.

A neural network~\cite{Feindt2006190} is used to further separate signal from background.
The network is trained using a simulation sample to represent signal and data from a $\Dstarm\Kp\pip$ mass sideband region to represent background.
The network exploits differences between signal and background in the distributions of 16 input variables related to the kinematics and topology of the decay.
The most discriminatory variables are the \Bp\ candidate $\chisqvtx$ and quantities related to the characteristic flight distances of the \Bp\ and \Dzb\ mesons.
It is verified that none of the input variables, nor the neural network output, are strongly correlated with the \Bp\ candidate mass or with position in the phase space of the \Bp\ meson decay.
The selection requirement on the neural network output is optimised using a figure of merit that does not depend on the assumed signal branching fraction~\cite{Punzi:2003bu}.
In this procedure, the relative efficiency of the neural network output requirement is determined from simulation, while the expected background under the signal peak in the \Bp\ candidate mass is obtained by extrapolating from a $\Dstarm\Kp\pip$ mass sideband region.
The same requirement is applied to both $\Dstarm\Kp\pip$ and $\Dstarm\pip\pip$ candidates.
The combined efficiency of the geometrical acceptance, online and offline selection (excluding particle identification) requirements is around $0.5\,\%$ for both $\Dstarm\Kp\pip$ and $\Dstarm\pip\pip$ final states.

Information from the ring-imaging Cherenkov detectors is combined with input from other subdetectors into variables designed to distinguish kaons from pions~\cite{LHCb-DP-2012-003}.
Requirements on the values of these variables for the pions and kaons originating directly from the \Bp\ decay and for the kaon from the \Dzb\ decay are imposed.
These are optimised using the same figure of merit as for the neural network output, with the signal efficiency determined from high-yield control samples of kaons and pions weighted to match the kinematic properties of signal decays.
Application of the same procedure to the pions from the \Dstarm\ and \Dzb\ decays indicates that no requirement is needed on the particle identification information associated with these particles.
The combined efficiency of the particle identification requirements is about $60\,\%$ for the $\Bp \to \Dstarm \Kp \pip$ decay and about $85\,\%$ for the $\Bp \to \Dstarm \pip\pip$ decay.
The $\pi \to K$ misidentification rate is below $0.3\,\%$.

Final-state particles from true $\Bz \to \Dstarm\pip$ decays can be combined with random pions to form a background which has a broad peak in the \Bp\ candidate mass above the signal region in the normalisation channel.  
In order to simplify the modelling of the background in this region, candidates with $\Dstarm\pip$ invariant mass in the range $5200 < m(\Dstarm\pip) < 5400 \mevcc$ are vetoed.
This requirement effectively removes the $\Bz \to \Dstarm\pip$ background with negligible loss of signal.
No veto is applied to remove the similar background from $\Bz \to \Dstarm\Kp$ decays as this is found to have negligible effect on the analysis.
Following all selection requirements, fewer than $2\,\%$ of events contain more  than one candidate; all are retained.
The associated systematic uncertainty is negligible.

Extended maximum likelihood fits to the distributions of candidates in \Bp\ candidate mass are used to determine the yields of $\Bp \to \Dstarm \Kp \pip$ and $\Bp \to \Dstarm \pip\pip$ decays.
The fits contain components to describe the signals, combinatorial background and partially reconstructed backgrounds.
The latter are decays of the type $\B \to \Dstarm h^+h^{\prime +}X$, where $X$ represents an additional particle that has not been included in the reconstructed decay chain.
The fit to the $\Bp \to \Dstarm \Kp \pip$ candidates also includes a component for cross-feed due to misidentified $\Bp \to \Dstarm \pip\pip$ decays.

The signal shapes are modelled by the sum of two Crystal Ball (CB) functions~\cite{Skwarnicki:1986xj}, which share a common peak position and have tails on opposite sides. 
The ratio of widths of the CB shapes and the fraction of entries in the narrower CB shape are constrained within their uncertainties to the values found in fits to simulated signal samples. 
The tail parameters of the CB shapes are fixed to those found in simulation. 
The combinatorial background in both samples is modelled with an exponential function.
Partially reconstructed background is modelled by the convolution of a Gaussian with an ARGUS function~\cite{Albrecht:1990am}, as this shape has been previously found to provide a good description of the kinematic limit for this component near $m_\B - m_\pi$~\cite{LHCb-PAPER-2016-004,LHCb-PAPER-2016-036}.
The cross-feed background is modelled with a CB function with parameters obtained from a fit to $\Bp \to \Dstarm \pip\pip$ data reconstructed with the kaon mass assigned to one of the daughters, weighted according to the misidentification probability obtained from control samples. 

\begin{figure}[!tb]
  \centering
  \includegraphics[width=0.49\textwidth,clip=true]{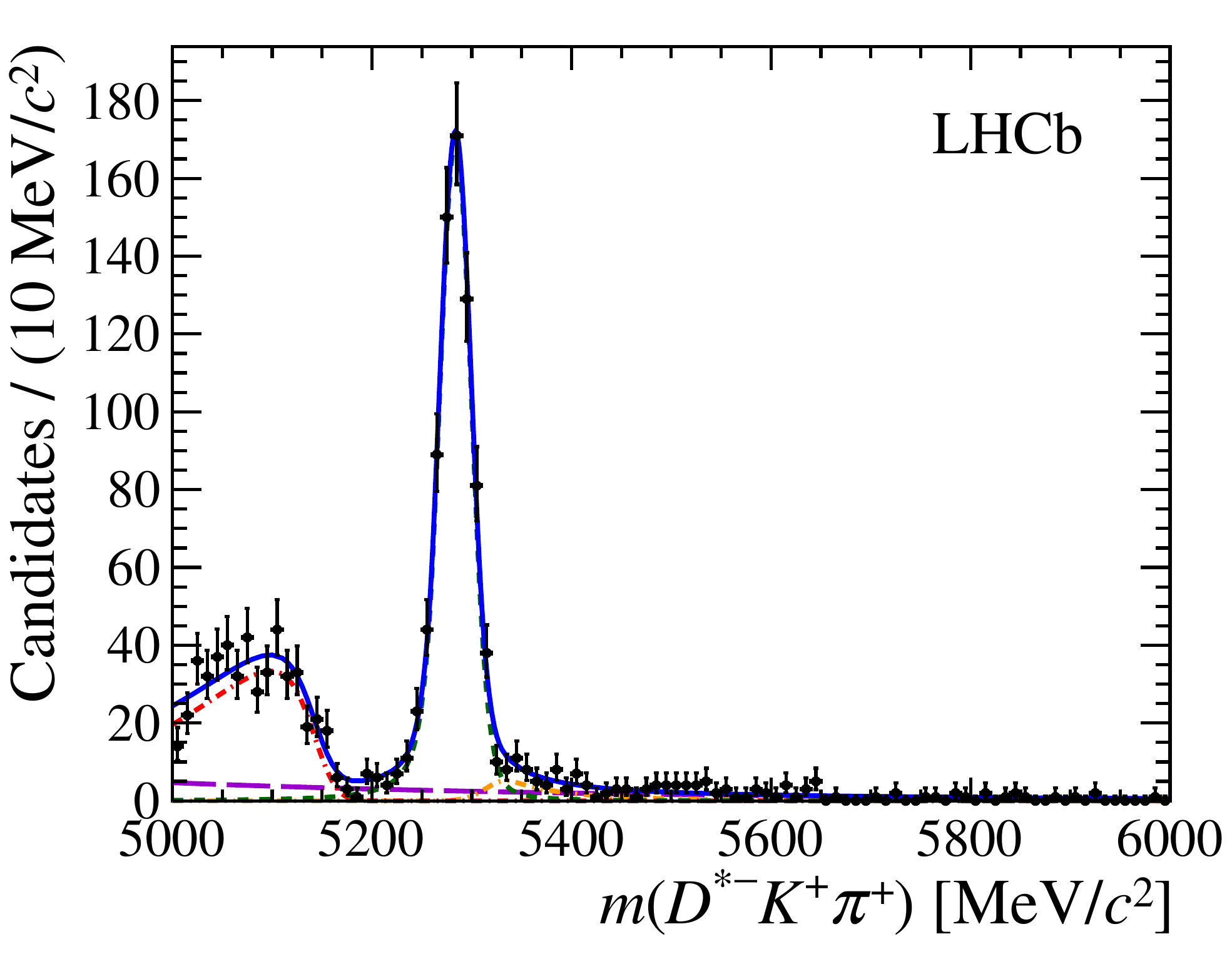}
  \includegraphics[width=0.49\textwidth,clip=true]{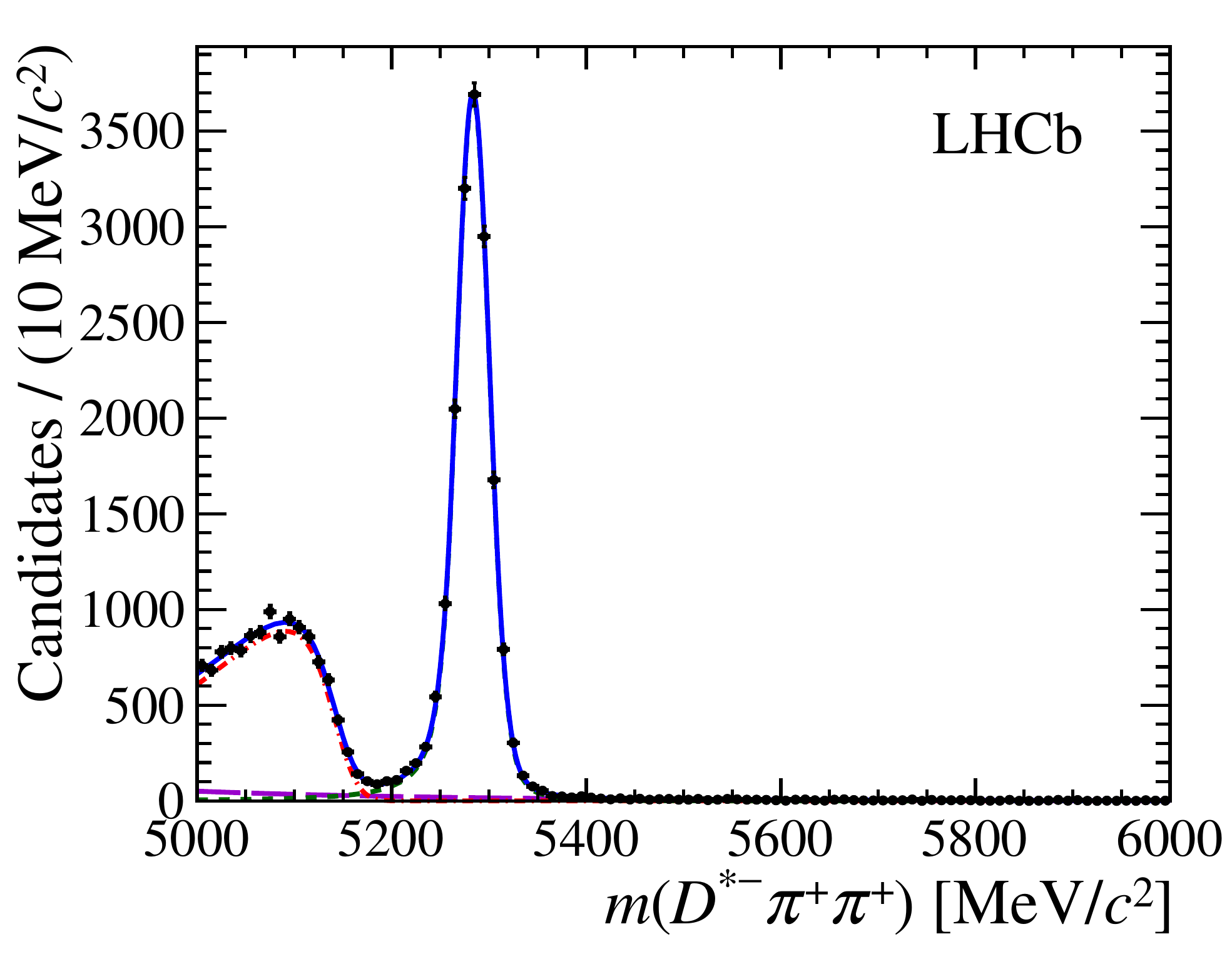}
  \includegraphics[width=0.49\textwidth,clip=true]{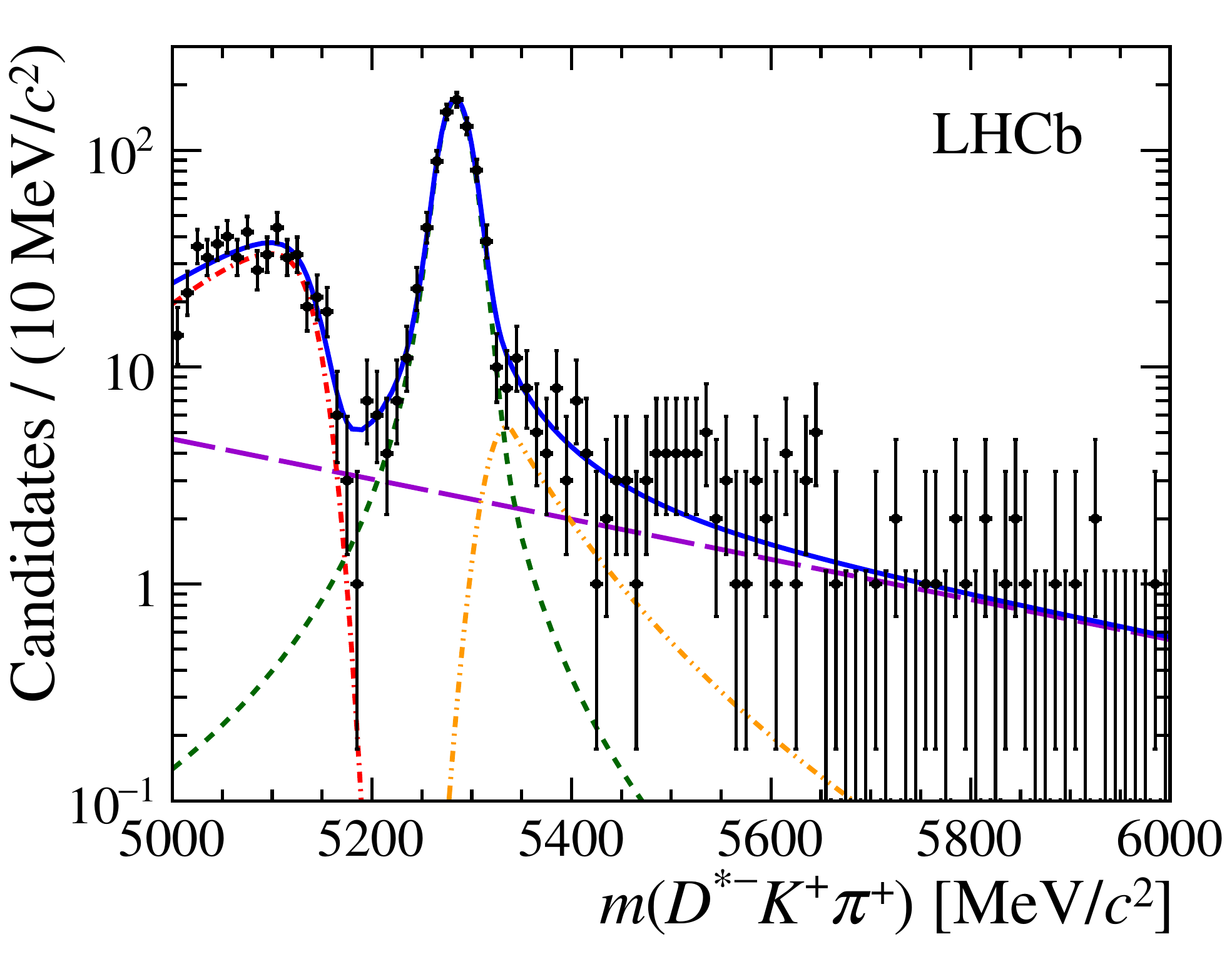}
  \includegraphics[width=0.49\textwidth,clip=true]{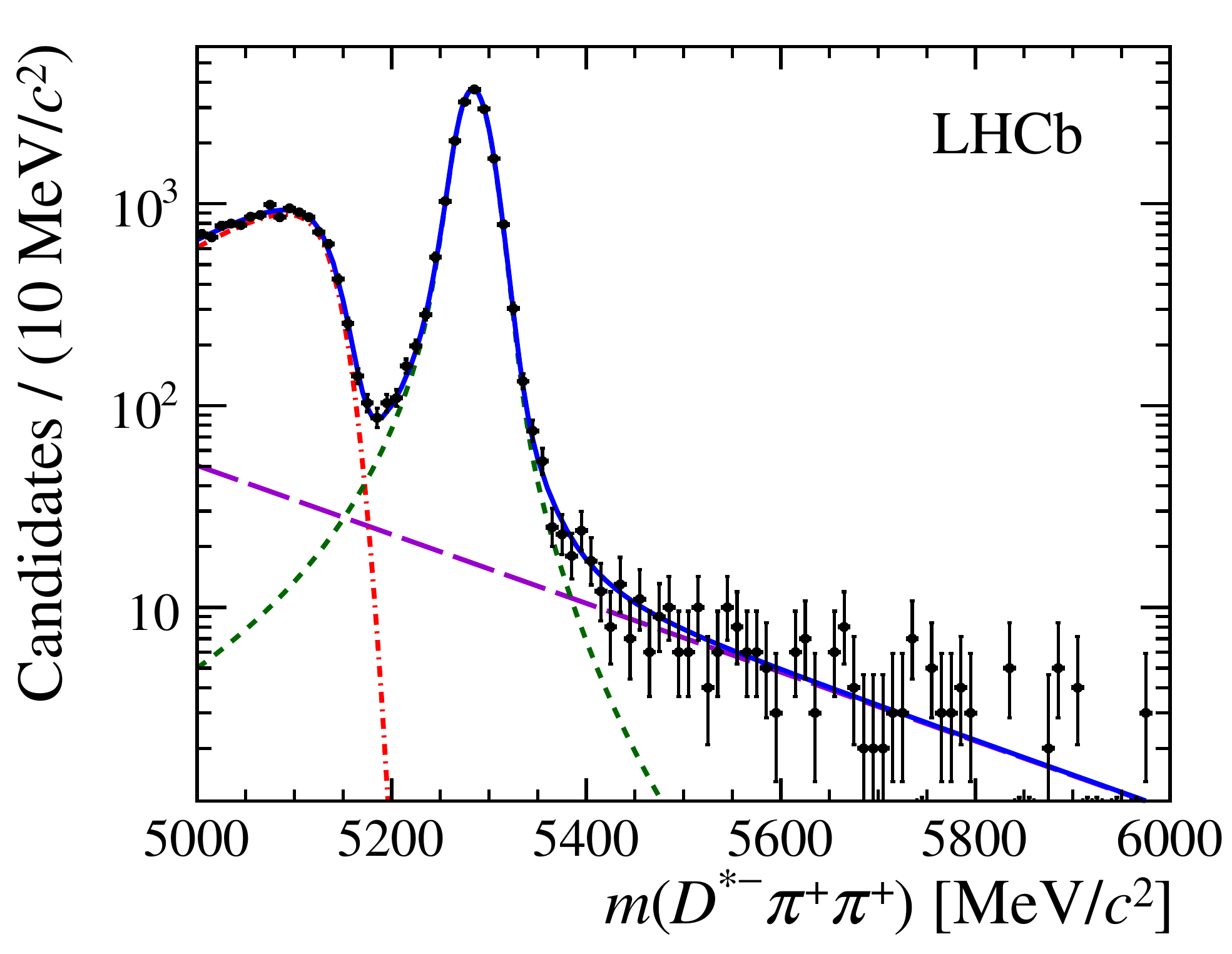}
  \caption{\small
    Fits to \B candidate mass distributions for (left) $\Dstarm \Kp \pip$ and (right) $\Dstarm \pip \pip$ samples with (top) linear and (bottom) logarithmic $y$-axis scales. 
    The individual components are (solid blue) total fit function, (dashed green) signal shape, (long-dashed violet) combinatorial background, (dot dashed red) partially reconstructed background and (double-dot dashed orange) $\Dstarm \pip \pip$ to $\Dstarm \Kp \pip$ cross-feed.
  }
  \label{fig:SigFit}
\end{figure}

The results of the fits are shown in Fig.~\ref{fig:SigFit}.
The fit to the $\Dstarm \Kp \pip$ sample has nine free parameters, which are the signal yield ($744 \pm 29$), the yields of the three background components, the peak position and width parameter of the signal shape, the slope of the combinatorial background and the two shape parameters of the partially reconstructed background.
The fit to the $\Dstarm \pip \pip$ sample has one fewer free parameter as no cross-feed component is included, and gives a signal yield of $17\,450 \pm 140$.
The fit procedure is validated with ensembles of pseudoexperiments; any possible bias on the fitted yields is found to be negligible.

The ratio of branching fractions for $\Bp \to \Dstarm \Kp \pip$ and $\Bp \to \Dstarm \pip\pip$ decays is calculated by applying event-by-event efficiency corrections as a function of position in the \Bp\ decay phase space,
\begin{equation}
  \label{eq:BFratio}
  \frac{{\cal B}(\Bp\to\Dstarm\Kp\pip)}{{\cal B}(\Bp\to\Dstarm\pip\pip)} = 
  \frac{N^{\rm corr}(\Bp\to\Dstarm\Kp\pip)}{N^{\rm corr}(\Bp\to\Dstarm\pip\pip)}\,,
\end{equation}
where $N^{\rm corr} = \sum_{i}W_i/\epsilon_i$ is the efficiency-corrected yield. 
Here the index $i$ runs over all candidates in the fitted data sample, $W_i$ is the signal weight for candidate $i$ and is determined using the \sPlot\ procedure~\cite{Pivk:2004ty} from the fits in Fig.~\ref{fig:SigFit}, and $\epsilon_i$ is the efficiency for candidate $i$.
The efficiencies are evaluated including contributions from the LHCb detector acceptance, selection and trigger. 
The acceptance and most selection efficiencies are calculated from simulated samples with, where appropriate, data-driven corrections applied, while the particle identification efficiency is determined from control samples~\cite{LHCb-DP-2012-003}.
The phase space for a $P \to VPP$ decay, where $V$\,($P$) indicates a vector (pseudoscalar) particle, has four degrees of freedom, but for $\Bp \to \Dstarm \Kp \pip$ it is found that the efficiency depends strongly only on the squares of the two-body invariant masses $m^2(\Dstarm\pip)$ and $m^2(\Kp\pip)$.
Similarly for $\Bp \to \Dstarm \pip \pip$ decays, dependence of the efficiency on $m^2(\Dstarm\pip)_{\rm min}$ and $m^2(\pip\pip)$ is accounted for, where $m^2(\Dstarm\pip)_{\rm min}$ indicates that the smaller of the two possible $m^2(\Dstarm\pip)$ combinations is taken.
The other two degrees of freedom in the phase space are related to the orientation of the $\Dstarm\to\Dzb\pim$ decay relative to the plane defined by the $\Bp \to \Dstarm h^+h^{\prime+}$ decay.
Possible variation of the efficiency with these variables is accounted for as a source of systematic uncertainty.

\begin{figure}[!tb]
  \centering
  \includegraphics[width=0.46\textwidth]{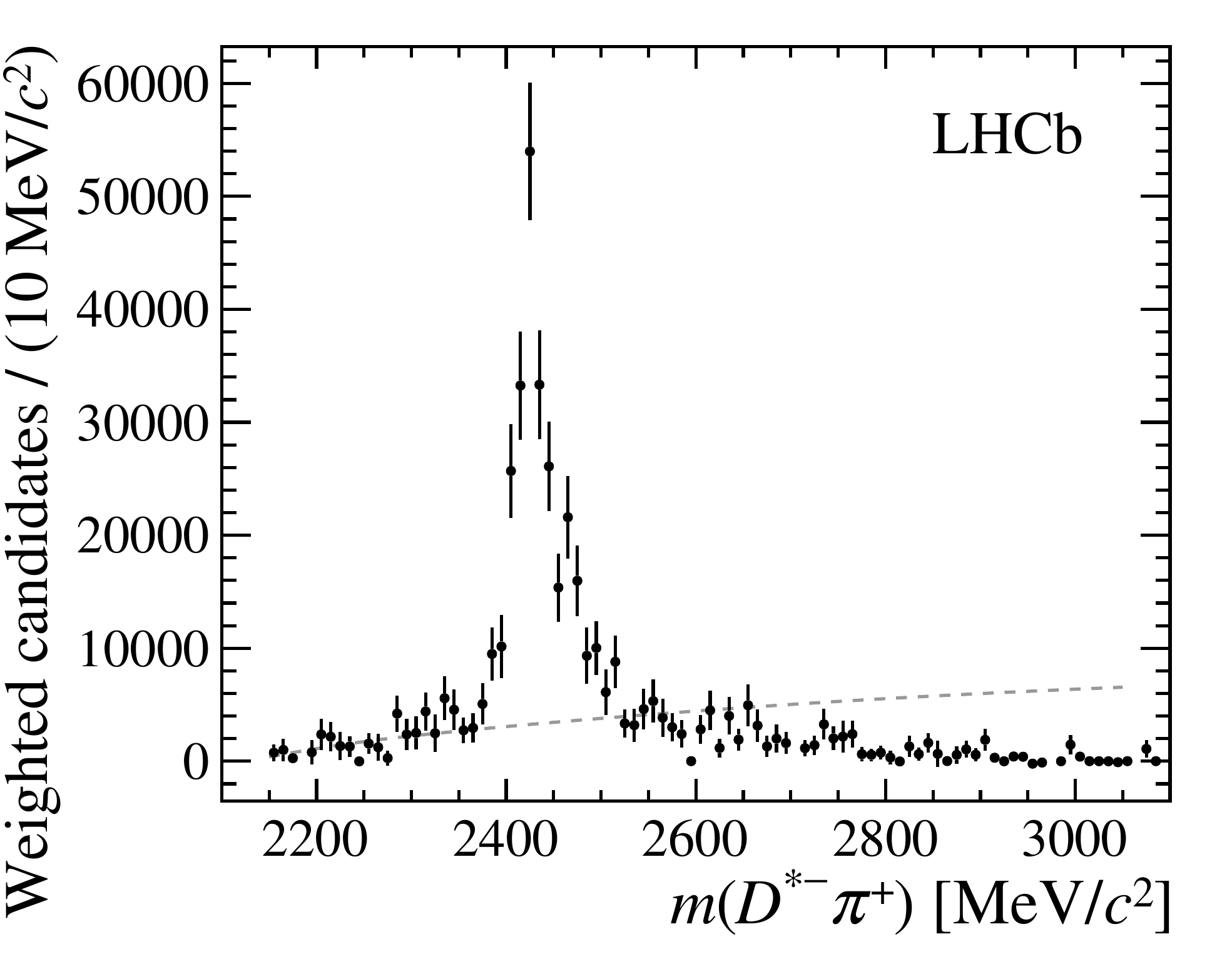}
  \caption{\small
    Background-subtracted~\cite{Pivk:2004ty} and efficiency-corrected $m(\Dstarm\pip)$ distribution from $\Bp\to\Dstarm\Kp\pip$ decays.
    The grey dashed line illustrates a phase-space distribution, normalised to the same number of weighted candidates.
  }
  \label{fig:1D-effcor}
\end{figure}

The background-subtraction and efficiency-correction procedures used to determine the values of $N^{\rm corr}$ also allow the phase-space distributions of decays to be examined.
The projection of the $\Dstarm\Kp\pip$ data onto $m(\Dstarm\pip)$ is shown in Fig.~\ref{fig:1D-effcor}.
The asymmetric peak is indicative of the presence of contributions from both the $\Dbar{}_1(2420)^0$ and $\Dbar{}_1^\prime(2430)^0$ states~\cite{Abe:2003zm}.
A detailed investigation of the distribution of decays across the phase space is left for future study.

The statistical uncertainty evaluated from Eq.~(\ref{eq:BFratio}) includes contributions from the weighting and from the floated shape parameters in the fit~\cite{LHCb-PAPER-2012-018}.
Systematic uncertainties are assigned due to approximations made in the fit used to determine the yields and due to uncertainties in the efficiency. 
Variations of the fit model are made by modifying fixed parameters within their uncertainties, replacing the shapes used to describe each component with alternative functions, and, in the fit to the $\Dstarm \pip\pip$ sample, introducing a component to account for cross-feed from $\Bp \to \Dstarm \Kp\pip$ decays.
Uncertainties on the efficiency arise due to the limited size of the simulation samples, possible variation of the efficiency with \Dstarm\ decay angles, possible imprecision of the data-driven method to determine particle identification efficiencies and due to selection requirements that remove particular regions of phase space.  
The magnitudes of each of these contributions are summarised in Table~\ref{tab:systSummary}.
The total systematic uncertainty on the ratio of branching fractions is $7.4\,\%$.

\begin{table}[tb]
\centering
\caption{
  Systematic uncertainties on the ratio of branching fractions.
}
\label{tab:systSummary}
\begin{tabular}{lc}
\hline
Source & Uncertainty \\
\hline
Fit model                              & $3.3\,\%$ \\
Simulation sample sizes                & $5.4\,\%$ \\
Efficiency variation with decay angles & $0.3\,\%$ \\
Particle identification efficiency     & $2.2\,\%$ \\
Phase space vetoes                     & $3.2\,\%$ \\
\hline
Total                                  & $7.4\,\%$ \\
\hline
\end{tabular}
\end{table}

The ratio of branching fractions is determined from Eq.~(\ref{eq:BFratio}) to be 
$$
\frac{{\cal B}(\Bp\to\Dstarm\Kp\pip)}{{\cal B}(\Bp\to\Dstarm\pip\pip)} = 
\left( 6.39 \pm 0.27 \pm 0.48 \right) \times 10^{-2} \, ,
$$
where the first uncertainty is statistical and the second is systematic.
This constitutes the first observation of the $\Bp\to\Dstarm\Kp\pip$ decay.
The change in $\sqrt{-2\ln {\cal L}}$ between fits with and without the signal component included, where ${\cal L}$ is the fit likelihood modified to account for systematic uncertainties that affect the yield, gives a value of 24, showing clearly that the significance is far in excess of the $5\,\sigma$ threshold normally used to claim observation.

In summary, the $\Bp\to\Dstarm\Kp\pip$ decay has been observed for the first time in a data sample corresponding to $3 \invfb$ of integrated luminosity recorded with the LHCb detector.
The ratio of the $\Bp\to\Dstarm\Kp\pip$  and $\Bp \to \Dstarm\pip\pip$ branching fractions has been measured, and has a value at the level na\"{i}vely expected due to the relative Cabibbo suppression of the former decay, $\left| V^{}_{us}/V^{}_{ud}\right|^2 \approx 5.3 \times 10^{-2}$.
The measurements that comprise the current world average value ${\cal B}(\Bp \to \Dstarm\pip\pip) = \left( 1.35 \pm 0.22 \right) \times 10^{-3}$~\cite{PDG2016,Abe:2003zm} all assume equal production of \Bp\Bm and \Bz\Bzb at the \FourS\ resonance.
Using this value and correcting it with the latest result on $\Gamma(\FourS\to\Bp\Bm)/\Gamma(\FourS\to\Bz\Bzb)$~\cite{PDG2016} results in
$$
{\cal B}(\Bp \to \Dstarm\Kp\pip) = 
\left( 8.2 \pm 0.3 \pm 0.6 \pm 1.3 \right) \times 10^{-5} \, ,
$$
where the third uncertainty is due to the precision of the knowledge of the normalisation channel branching fraction.
Inspection of the phase-space distribution of signal decays confirms that this mode can be used to investigate charm meson spectroscopy.

%% file: acknowledgements.tex
\section*{Acknowledgements}

\noindent We express our gratitude to our colleagues in the CERN
accelerator departments for the excellent performance of the LHC. We
thank the technical and administrative staff at the LHCb
institutes. We acknowledge support from CERN and from the national
agencies: CAPES, CNPq, FAPERJ and FINEP (Brazil); MOST and NSFC (China);
CNRS/IN2P3 (France); BMBF, DFG and MPG (Germany); INFN (Italy); 
NWO (The Netherlands); MNiSW and NCN (Poland); MEN/IFA (Romania); 
MinES and FASO (Russia); MinECo (Spain); SNSF and SER (Switzerland); 
NASU (Ukraine); STFC (United Kingdom); NSF (USA).
We acknowledge the computing resources that are provided by CERN, IN2P3 (France), KIT and DESY (Germany), INFN (Italy), SURF (The Netherlands), PIC (Spain), GridPP (United Kingdom), RRCKI and Yandex LLC (Russia), CSCS (Switzerland), IFIN-HH (Romania), CBPF (Brazil), PL-GRID (Poland) and OSC (USA). We are indebted to the communities behind the multiple open 
source software packages on which we depend.
Individual groups or members have received support from AvH Foundation (Germany),
EPLANET, Marie Sk\l{}odowska-Curie Actions and ERC (European Union), 
Conseil G\'{e}n\'{e}ral de Haute-Savoie, Labex ENIGMASS and OCEVU, 
R\'{e}gion Auvergne (France), RFBR and Yandex LLC (Russia), GVA, XuntaGal and GENCAT (Spain), Herchel Smith Fund, The Royal Society, Royal Commission for the Exhibition of 1851 and the Leverhulme Trust (United Kingdom).

%% file: LHCb_Authorship_flat_16-Feb-2017.tex
\centerline{\large\bf LHCb collaboration}
\begin{flushleft}
\small
R.~Aaij$^{40}$,
B.~Adeva$^{39}$,
M.~Adinolfi$^{48}$,
Z.~Ajaltouni$^{5}$,
S.~Akar$^{59}$,
J.~Albrecht$^{10}$,
F.~Alessio$^{40}$,
M.~Alexander$^{53}$,
S.~Ali$^{43}$,
G.~Alkhazov$^{31}$
% ...

{\footnotesize \it
% ...
$ ^{5}$Clermont Universit{\'e}, Universit{\'e} Blaise Pascal, CNRS/IN2P3, LPC, Clermont-Ferrand, France\\
% ...
$ ^{10}$Fakult{\"a}t Physik, Technische Universit{\"a}t Dortmund, Dortmund, Germany\\
% ...
$ ^{31}$Petersburg Nuclear Physics Institute (PNPI), Gatchina, Russia\\
% ...
$ ^{39}$Universidad de Santiago de Compostela, Santiago de Compostela, Spain\\
$ ^{40}$European Organization for Nuclear Research (CERN), Geneva, Switzerland\\
% ...
$ ^{43}$Nikhef National Institute for Subatomic Physics, Amsterdam, The Netherlands\\
% ...
$ ^{48}$H.H. Wills Physics Laboratory, University of Bristol, Bristol, United Kingdom\\
% ...
$ ^{53}$School of Physics and Astronomy, University of Glasgow, Glasgow, United Kingdom\\
% ...
$ ^{59}$University of Cincinnati, Cincinnati, OH, United States\\
% ...
}
\end{flushleft}